\documentstyle[aps,prb,epsfig,floats]{revtex}

\def\cdcrse{CdCr$_2$Se$_4$}

\begin{document}

\wideabs{ 
\title{First-principles study of band offsets 
in ferromagnetic semiconductor heterojunctions}
\author{James M. Sullivan and Steven C. Erwin}
\address{Center for Computational Materials Science,
Naval Research Laboratory, Washington, D.C. 20375}
\date{\today}
\maketitle
\begin{abstract}
We report valence and conduction band alignments and offsets for heterojunctions
between CdCr$_2$Se$_4$, an {\it n}-type ferromagnetic semiconductor, and the non-magnetic
materials Si and GaAs, evaluated using density functional theory. We explore numerically 
the impact of different interface features on the type of band alignment
and the magnitude of the offsets. For example, we find it is energetically favorable to
deplete Cr atoms from the layers at the interface; this also leads to
band alignments smaller in magnitude compared to those obtained for Cr-rich interfaces 
and ideal for electrical spin-injection into either Si or Ga-terminated GaAs substrates.
\end{abstract}
}

\section{Introduction}\label{Introduction}

Spin ``injection'' refers to the induced transport of spin-polarized
electrons or holes from one material into another. To date, two basic
approaches to injection have been explored: photoinduced generation of
spin-polarized carriers in one material, which then diffuse across an
interface into a second material;
\cite{koshihara97,malajovich99,malajovich01} and direct electrical
injection via application of a bias voltage across an
interface.\cite{fiederling99,ohno99,jonker00,filip00} In both
approaches, the heterojunction band offset represents an important
quantity which may place fundamental constraints on the expected
efficiency of the injection process.  In this work we report a
first-principles numerical study of band offsets for realistic
heterojunctions between magnetic and nonmagnetic semiconductors.

We will refer to the magnetic and nonmagnetic sides of the
heterojunction as the ``source'' and ``sink,'' respectively. Our
choice of the source and sink materials are guided by a few simple
principles. First, since spin lifetimes are generally longer for 
electrons\cite{eref} and since electrons generally have higher mobilities 
and smaller spin-orbit interaction than holes,\cite{harrison80} we focus 
on source materials which can supply spin-polarized electrons rather than holes.  
Second, we restrict our attention to semiconducting source materials so as to 
avoid the well-known ``conductivity mismatch'' problem that plagues direct injection 
from a metal into a semiconductor.\cite{schmidt00} Third, because lattice
mismatch leads to dislocations (which act as spin scatterers), we
consider sink materials with a reasonably close lattice match to the
source material.  A particularly interesting candidate source material
is the chromium chalcogenide spinel \cdcrse, 
an {\it n}-type ferromagnetic semiconductor with a relatively high
Curie temperature of 130 K.\cite{lehmann67} For sink materials, we
consider Si and GaAs, both of which can be made {\it n}-type with
appropriate doping and are well lattice matched to \cdcrse.

\begin{figure}[tb]
\centerline{\epsfig{figure=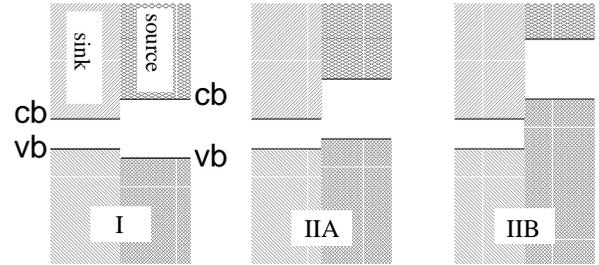,width=8.0cm}}
\caption{ \label{bandalignments_fig} Various types
of band alignments in a semiconducting heterojunction. Valence and
conduction band edges are denoted vb and cb, respectively.  }
\end{figure}

In this work we use density functional theory to calculate band
alignments and offsets between \cdcrse\ and Si or
GaAs. We take the view that although the band alignment may ultimately
place constraints on the injection efficiency, one may be able to
``tailor'' the interface during the growth process, thereby improving
alignments that are otherwise unfavorable for injection. It is this
variability in the alignments and offsets, with respect to various features
of the interface, that we explore numerically in this work.
Of the three possible types of alignment (see
Fig.~\ref{bandalignments_fig}), the two most favorable for 
spin-injection of electrons are Type I or IIA.  Less favorable is Type 
IIB alignment, which leads to Zener breakdown \cite{zener_ref} (valence
electrons of the source material tunnel into the conduction bands of
the sink). Type IIB alignment also gives rise, through band bending, to
confinement of holes in the source material in the layers adjacent to
the interface, which would then act as strong spin-scattering
centers and presumably degrade spin transport across the interface.

\section{Theory of band offsets}\label{Theory}

In this work we evaluate the band offsets using methods pioneered by
Van de Walle and
Martin,\cite{vandewalle86,vandewalle87,vandewalle89,franciosi96,qteish92}
here extended to include spin splitting. For
completeness we review the essentials of that approach. We take
the sink material to be on the left (L) side of the heterostructure,
and the source material to be on the right (R).
The valence band offset (VBO)
is the relative alignment of the L and R valence band maxima (VBM) far from
the interface, and is conveniently written as
\begin{equation}
E_{\rm VBO}= \Delta E_{\rm band} + \Delta V_{\rm avg}.\label{vbo_eq}
\end{equation}
Here $\Delta E_{\rm band}$ and $\Delta V_{\rm avg}$ are generally
referred to as the {\it band lineup} and {\it potential lineup},
respectively.  $\Delta E_{\rm band}$ is determined entirely by the
electronic structure of the constituent {\it bulk} materials, while
$\Delta V_{\rm avg}$ accounts for the effects of the {\it interface}
between those materials.  For simplicity we do not show spin indices
in Eq.~\ref{vbo_eq}; each spin channel is treated independently.

\begin{figure}[tb]
\centerline{\epsfig{figure=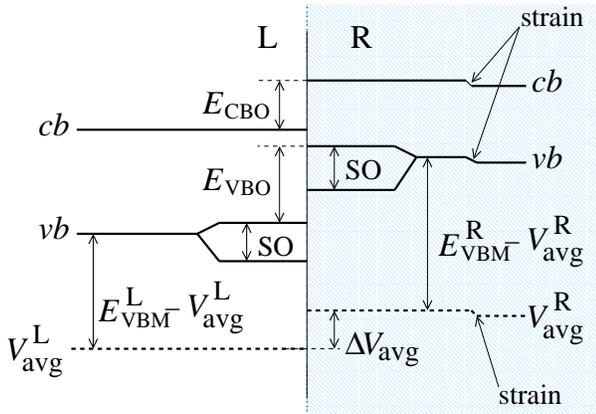,width=8.0cm}}
\caption{ \label{bandlevels_fig} The various contributions to 
the valence band offset, $E_{\rm VBO}$. 
$V_{\rm avg}$ is the average value of the effective potential in L
and R far from the interface, to which the valence band maxima
are referenced. Also shown are the level shifts arising from strain 
in the overlayer (R), as well as spin-orbit (SO) splitting in L and R.  
Only one spin channel is depicted.}
\end{figure}

Although the band lineup, $\Delta E_{\rm band}$, is a purely bulk
quantity, its evaluation in general depends on the nature of the
interface.  For example, lattice mismatch between the substrate and
overlayer will introduce tetragonal strain into an overlayer that has
a nominally cubic lattice.  This strain will in general lead to level
shifts and band splittings that must be correctly included when
computing the bulk band lineup.  For the purposes of this work we
consider the nonmagnetic sink materials to be unstrained substrates on
which overlayers of \cdcrse\ are grown epitaxially. Thus the overlayers
experience an in-plane tensile strain (1.3\% and 3.3\% for Si and GaAs,
respectively) that results in a contraction of the spacing between the
overlayers. To fully account for these strain effects we consider bulk
unit cells for \cdcrse\ with a tetragonal distortion 
$a_{\rm 1} = a_{\rm 2} = 2a_{\rm substrate} \neq a_{\rm 3}$.

We also include fully the effects of the spin-orbit (SO) interaction on the
band structure, by performing fully relativistic calculations for the
bulk phases.  With strain and spin-orbit effects thus included,
the band lineup takes the form
\begin{eqnarray}
\Delta E_{\rm band} &=& (E_{\rm VBM}^{\rm L} - V_{\rm avg}^{\rm L} + 
\Delta E_{\rm SO}^{\rm L})\label{eb_eq}  
\nonumber\\
&-& (E_{\rm VBM}^{\rm R} - V_{\rm avg}^{\rm R} + \Delta E_{\rm SO}^{\rm R}).
\end{eqnarray}
Here, $E_{\rm VBM}$ is the position of the valence-band maximum, and
$V_{\rm avg}$ is the cell-averaged Kohn-Sham potential (which provides
a convenient reference energy). $\Delta E_{\rm SO}$ is the shift in the VBM 
associated with spin-orbit interaction. Figure~\ref{bandlevels_fig} shows an an 
energy level-diagram including all the various terms defined above.

The second term in Eq.~\ref{vbo_eq}, the potential lineup $\Delta
V_{\rm avg}$, is determined by the electronic structure of a specific
interface. We use periodic supercells to represent the heterojunction,
with supercell periods sufficiently large to ensure that the
interfaces do not interact.  The potential lineup is the difference in
the locally averaged potential on the two sides of the interface, far from
the interface itself. To compute these local averages we use the
slab-averaging technique discussed in Ref.\onlinecite{franciosi96}. 
One first computes the planar average,
$\overline{V}(z)$, of the total potential, $V({\bf r})$, by averaging across
planes parallel to the interface.  $\overline{V}(z)$ contains atomic-scale
variations whose local periodicity on the two sides of the interface
may in general may be different---even far from the
interface---because of lattice mismatch. The correct local averages can be
obtained by performing a ``slab average'' over a sliding window. This slab
averaging is applied twice, in order to allow for different L and R windows:\cite{franciosi96,allak01}
\begin{equation}
\overline{\overline{V}}(z) \equiv \frac{1}{\lambda_{L}\lambda_{R}}\ 
\int_{z-\lambda_{L}/{2}}^{z+\lambda_{L}/2}
\int_{z''-\lambda_{R}/2}^{z''+\lambda_{R}/2} \overline{V}(z') dz'dz''. 
\label{slab_eq}
\end{equation}
Here $\lambda_{L}$ and $\lambda_{R}$ denote the periodic repeat
distances in the direction perpendicular to the interface for the L
and R materials . Far from the interface, the slab-averaged potential,
$\overline{\overline{V}}(z)$, is constant. The potential lineup is then given by
\begin{equation}
\Delta V_{\rm avg} = \overline{\overline{V}}(z_{L}) - \overline{\overline{V}}(z_{R}), \label{avgpot_eq}
\end{equation}
where $z_{L}$ and $z_{R}$ denote the midpoints of the L and R slabs.
Finally, having computed the VBO directly from Eq.~\ref{vbo_eq}, the 
conduction band offset (CBO) is simply 
\begin{equation}
E_{\rm CBO}= E_{\rm VBO} + E_{\rm gap}^{L}-E_{\rm gap}^R, \label{cbo_eq}
\end{equation}
where $E_{\rm gap}^{L(R)}$ is the band gap for L (R). For reasons discussed
below (see Sec.~\ref{bulk_cdcrse_bands}), we evaluate Eq.~\ref{cbo_eq} using experimental
band gaps rather than our theoretical values.

\section{Interface Models}\label{InterfaceModels}

\begin{figure}[tb]
\centerline{\epsfig{figure=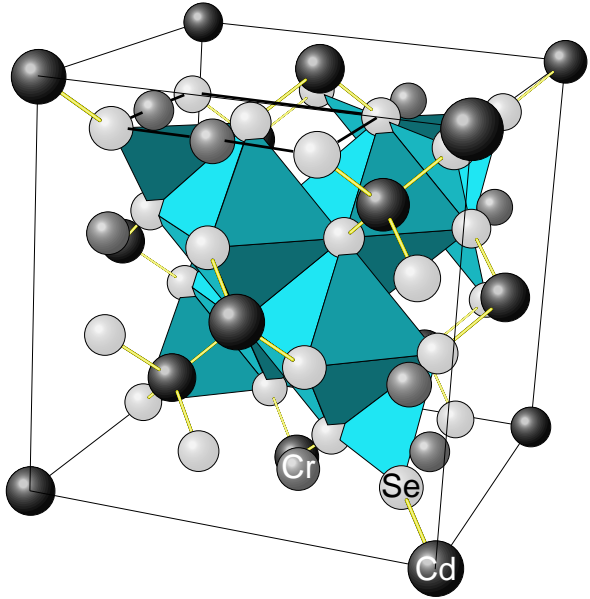,width=8.0cm}}
\caption{ Crystal structure of \cdcrse\ in the bulk spinel
phase. Representative Cd, Cr, and Se atoms are labeled. The Se atoms
form the vertices of octahedra (some of which are shown) centered at
each Cr site. A (001) plane containing Cr and Se atoms is also indicated.
\label{structure_fig} }
\end{figure}

The crystal structure of bulk \cdcrse\ is shown in
Fig.~\ref{structure_fig}. In this work we consider only (001)
interfaces, for which tetrahedral bonding across the interface can be
maintained in a particularly simple manner. In
Fig.~\ref{structure_fig}, different (001) basal planes can formed by
truncating the spinel structure at different atomic planes. Only two 
chemically inequivalent planes can be so formed: one containing only
Cd atoms, and one containing Se and Cr atoms in a 2:1 ratio.  Such a
representative plane containing Se and Cr atoms is indicated in Fig.~\ref{structure_fig}. 
Within these planes, the atomic density in the Cd plane 
is only one fourth that of Si or GaAs (ignoring the lattice mismatch), 
which suggests that a 2$\times$2 reconstruction of the terminating Si layer 
would result for interfaces with the Cd plane. On the other hand, for interfaces 
with the mixed (Se,Cr) plane a simple unreconstructed interface
is quite plausible, and so we limit ourselves to variants of this
basic interface geometry.

\begin{figure}[bt]
\centerline{\epsfig{figure=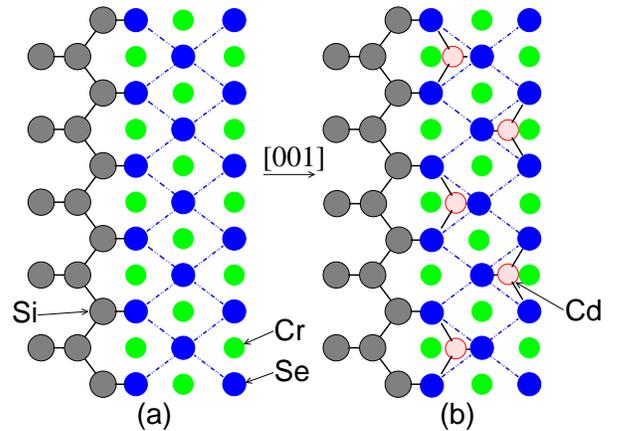,width=8.0cm}}
\caption{ Schematic of an abrupt [001] interface between \cdcrse\ 
and Si. Representative Cd, Cr, Se and Si atoms are  
denoted, as are the outlines of the octahedron defined by the Se atoms. Some aspects
of this figure (bond lengths, angles, etc.) are not drawn to 
scale. Note that not all atoms in this figure lie in the same plane. 
\label{interface1_fig}
}
\end{figure}

For interfaces with Si, the atomic alignment at the interface is shown
schematically in Fig.~\ref{interface1_fig}.  For the case of Si we
consider only atomically abrupt interfaces.  At the interface
boundary, every Si atom is tetrahedrally coordinated (to two Si and
two Se atoms). The role of Cr atoms at the
interface is not immediately obvious from these preliminary
considerations, but will be analyzed in detail in
Sect.~\ref{results}.

\begin{figure}[bt]
\centerline{\epsfig{figure=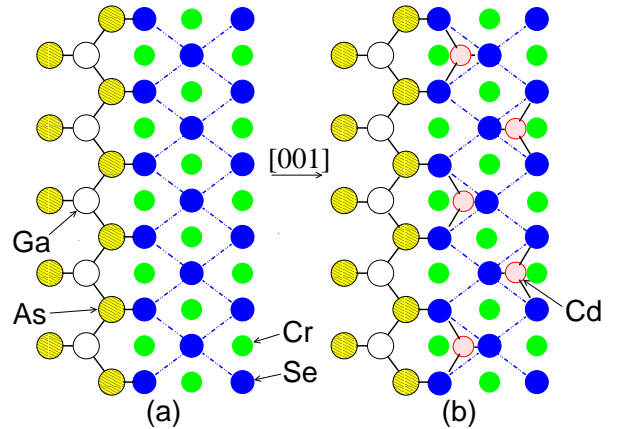,width=8.0cm}}
\caption{ Schematic of an abrupt [001] interface between \cdcrse\ 
and GaAs.
\label{interface2_fig} 
}
\end{figure}

For interfaces with GaAs we consider both abrupt and partially
intermixed interfaces.  In both cases we compute offsets for both
Ga-terminated and As-terminated
substrates. Fig.~\ref{interface2_fig} shows an abrupt As-terminated
substrate; Ga-terminated substrates are analogous.

\begin{figure}[bt]
\centerline{\epsfig{figure=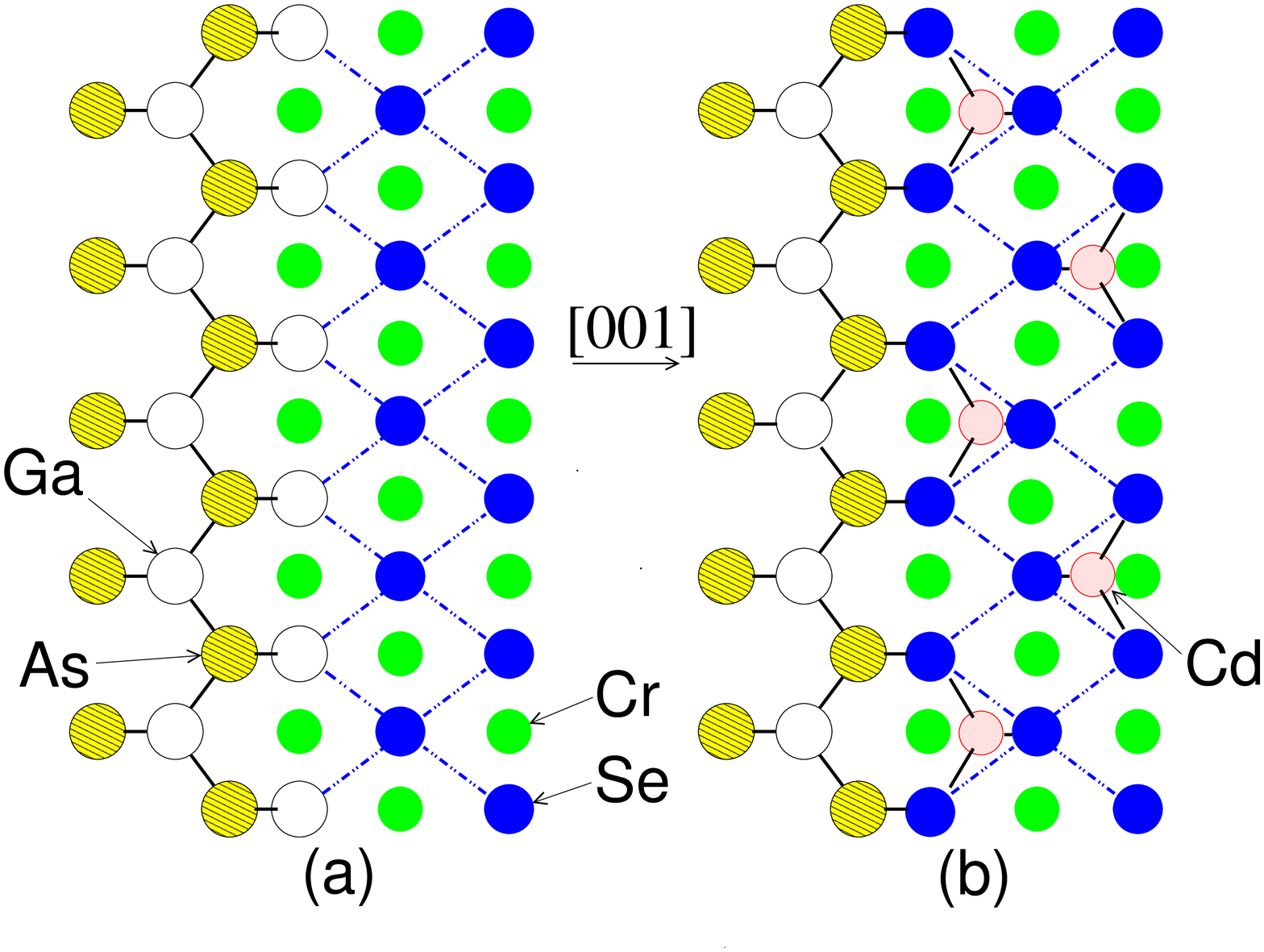,width=8.0cm}}
\caption{ Schematic of partially intermixed [001] interface between \cdcrse\ 
and GaAs. From simple counting arguments As-Se bonds contribute an
extra $\frac{3}{4}$ electron while Ga-Cd bonds contribute an excess $\frac{3}{4}$
hole. Thus substitution of Ga for Se in the interface layer of plane
(a) [cf. Fig.~\ref{interface2_fig}(a)] maintains local charge neutrality
over the first few interface layers.
\label{interface3_fig}  
}
\end{figure}

Although abrupt interfaces are conceptually attractive, for a polar
material like GaAs they may not be physically plausible. In
particular, for interfaces between polar and non-polar materials,
abrupt termination leads inevitably to the formation of long-range
dipole fields.\cite{harrison78} Such a dipole field is energetically
unfavorable over large distances, and will typically lead to an atomic
rearrangement near the interface in order to reduce such effects. For
polar/nonpolar interfaces, a particularly simple way of eliminating
the dipole field is to intermix the polar and nonpolar constituents at
the interfacial layer in order to maintain local charge
neutrality.\cite{franciosi96} We have followed this approach for
interfaces with GaAs, by appropriately substituting Ga or As atoms for
Se atoms at the interface in such a way as to maintain
local neutrality. Fig.~\ref{interface3_fig} shows an example of this
substitution for an As-terminated interface.

\section{Computational Details}\label{Num_Details}
For both the interface and bulk calculations we use the plane-wave
pseudopotential method with norm-conserving
pseudopotentials,\cite{troullier91} as implemented in the {\sc abinit} code.
\cite{abinit} In both methods we use the local-spin-density
approximation (LSDA) in the form of Perdew and Wang.\cite{perdew92}
We have carefully checked the accuracy of the pseudopotentials
by comparing to results for bulk \cdcrse\ obtained
with the linearized augmented plane-wave (LAPW) method as implemented
in the {\sc wien97} code.\cite{wien} Table~\ref{tab_opt} summarizes
the predictions of both methods for equilibrium lattice constant, bulk
modulus, and magnetic moment, along with the available experimental
values. The pseudopotential and LAPW lattice constants agree to better
than 1\%, the bulk moduli to better than 3\%, and the magnetic moment
essentially exactly.  The agreement with experimental values is
equally good with the exception of the bulk modulus (for which the
spread in experimental values is very large). 
We have also computed
the difference in total energy between the normal- and inverse-spinel
forms of \cdcrse,
\begin{equation}
\Delta E_{s} = E_{\rm normal} - E_{\rm inverse}, \label{struc_energy_eq}
\end{equation}
and find that the spinel structure is energetically preferred 
(see Table~\ref{tab_opt}) within both
PSPW and LAPW methods. Thus for the rest of this work we
consider only the normal spinel phase.

\begin{table} [bt]
\caption{\label{tab_opt}
Structural and magnetic properties of bulk \cdcrse,
as predicted by the pseudopotential plane-wave and LAPW methods
and as measured experimentally. The four columns are lattice constant,
bulk modulus, magnetic moment, and normal- vs. inverse-spinel total
energy.}
\begin{tabular}{lcccc}
Method    & $a$ (\AA) & $B$ (GPa)     &  $M$ ($\mu_{\rm B}$) & $\Delta E_{\rm s}$ (eV) \\
\hline
 PSPW    & \dec 10.68    & \dec 96.2   & \dec 6.0          & \dec $-$2.72 \\
 LAPW    & \dec 10.59    & \dec 93.9   & \dec 6.0          & \dec $-$1.97  \\
 Exp.   & \dec 10.72$^{\rm a}$ & 76.9$^{\rm b}$, 43.4$^{\rm c}$ & 5.4$^{\rm b}$, 5.6$^{\rm d}$ & \_\ \\
\end{tabular} 
$^{\rm a}$Ref. \onlinecite{baltzer65} \\
$^{\rm b}$Ref. \onlinecite{batlogg78} \\
$^{\rm c}$Ref. \onlinecite{galdikas89} \\
$^{\rm d}$Ref. \onlinecite{baltzer66} \\
\end{table}
The interface results we report here were obtained with supercells containing
five atomic layers of Si or GaAs and five layers of \cdcrse. These
thicknesses are sufficient to ensure that the interfaces do not interact, 
as confirmed by the convergence (discussed below) of the planar-averaged potentials to 
their bulk values within the first few layers of the interface. The coordinates of all 
the atoms were completely relaxed until all force components were smaller than
$10^{-3}$ Ha/bohr. By simultaneously optimizing the supercell
dimension normal to the interface, we impose no artificial constraints
on the interlayer spacing throughout the supercell. As described
above, however, we do constrain the in-plane lattice constant to that
of Si or GaAs. The resulting tensile strain experienced by the
\cdcrse\ overlayers gives rise to a decrease in the spacing between
bulk-like \cdcrse\ layers, consistent with total energy calculations
for the bulk material with a tetragonal distortion. The 
cumulative effect of all these physical relaxations (relative to the ideal geometry) 
on the absolute band offsets is typically of the order of 0.1--0.3 eV, 
and so must be fully included for an accurate description of the alignments.

For all the pseudopotential calculations, a plane-wave cutoff of 70 Ry
was used, which is sufficient to converge all band states relevant to
the offset calculations to within 100 meV. A 4$\times$4 Monkhorst-Pack \cite{monkhorst76}
sampling of the surface Brilloun zone was used for the supercell calculations,
equivalent to 8$\times$8 sampling of the primitive Si or GaAs surface
zone.

\section{Bulk \cdcrse\ Band Structure}\label{bulk_cdcrse_bands}
The electronic structure of bulk \cdcrse\ has been the subject of a
number of experimental studies.  Photoluminescence experiments give
the {\it fundamental gap} ---the energy difference between the highest
{\it p}-like valence state and lowest {\it s}-like conduction
state---to be 1.8 eV in the ferromagnetic phase.\cite{yao81}   In
contrast, optical studies reveal an absorption edge only 1.3 eV above 
the valence-band edge, suggesting the existence of {\it d}-like conduction 
states in the fundamental gap.\cite{batlogg78} This view is supported by
transport measurements showing relatively low mobility, consistent with
conduction taking place within narrow {\it d}-like bands. 
\cite{coutinho-filho79}

\begin{figure}[bt]
\centerline{\epsfig{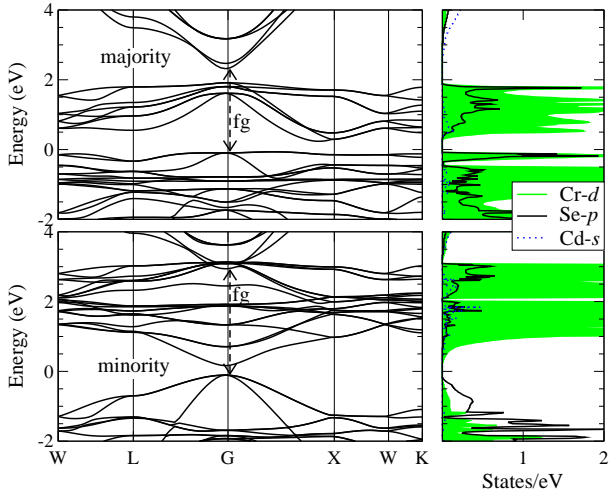}}
\caption{ Electronic structure of bulk CdCr$_2$Se$_4$ in the ferromagnetic
phase for the majority (upper panels) and minority (lower panels)
spin channels. The arrows labeled ``fg'' marks the {\it s-p} fundamental gap.
The valence band maximum is taken to be the energy zero.
\label{cdcrse_bands_fig}  
}
\end{figure}

Our results for the band structure of bulk \cdcrse\ are show in
Fig.~\ref{cdcrse_bands_fig}. We find that the ground state is indeed a
ferromagnetic semiconductor, with a magnetic moment of 6.0 $\mu_{B}$
per formula unit.  Our results are in very good agreement with other
recent LAPW calculations,\cite{continenza94} but are slightly
different from results obtained with the linearized-muffin-tin-orbital
(LMTO) method, which gives semimetallic band structure in both spin
channels and a slightly higher moment of 6.2 $\mu_{\rm B}$ per formula
unit.\cite{shanthi00}

Figure~\ref{cdcrse_bands_fig} shows the density of states for bulk \cdcrse\ decomposed
into angular character and projected onto atomic sites. Each Cr site contributes
a moment of ~$3 \mu_{B}$ to the total moment, corresponding to four spin-up and one
spin-down electron. Our predicted fundamental gaps, indicated in Fig.~\ref{cdcrse_bands_fig}
by arrows, are found to be 2.42 and 3.05 eV for the 
majority and minority spin channels respectively; these results are in very good 
agreement with those reported in Ref.\onlinecite{continenza94}. 

For spin injection, the relevant energy gap is the band gap between
valence and conductions states.  Our results predict the band gap to
be very small, of order 0.3 eV in both spin channels; this value is
much smaller than the experimental optical absorption edge at 1.3 eV.
Since there is experimental evidence suggesting that conduction takes
place in the unoccupied {\it d} levels probed by optical
absorption,\cite{coutinho-filho79,lederer72} we simply adopt the
experimental value of 1.3 eV for the band gap of \cdcrse\ appearing in 
Eq.~\ref{cbo_eq}. For similar reasons, we use experimental values of the band gaps of 
Si and GaAs.\cite{sigaasgap}

\section{Calculated Band Offsets}\label{results}
\begin{figure}[bt]
\centerline{\epsfig{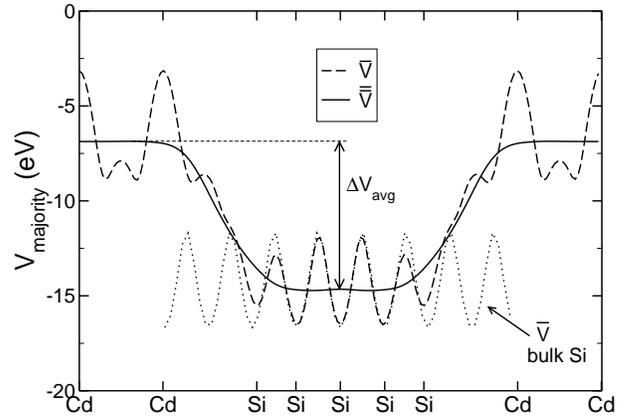}}
\caption{ Planar (dashed curve) and slab-averaged (solid curve) potentials 
in a CdCr$_2$Se$_4$/Si heterojunction for the majority spin channel (minority
spin channel is similar). The potential lineup is also indicated. The planar-averaged 
potential for bulk Si is shown for reference (dotted curve).
\label{si1dproj_fig} 
}
\end{figure}

We begin this section by illustrating the slab-averaging procedure
used to determine the potential lineup across an
interface. Figure~\ref{si1dproj_fig} shows the planar-averaged
and slab-averaged potentials for the case of
\cdcrse/Si. The five layers of Si are in the middle portion of the
figure, the atomic planes corresponding to minima in the planar-averaged
potential; the three maxima in the \cdcrse\ portion of the
cell correspond to Cd planes. The slab-averaged potential is quite
flat at the midpoints of the Si and \cdcrse\ slabs, demonstrating that
these slabs are sufficiently thick to converge the
potential-lineup contribution to the band offset.
Figure~\ref{si1dproj_fig} also shows the planar-averaged
potential for bulk Si. The rapid convergence of the planar-averaged
potential in the supercell to its bulk value is readily apparent, and
is similar to that found in previous studies. \cite{franciosi96}

\begin{figure}[bt]
\centerline{\epsfig{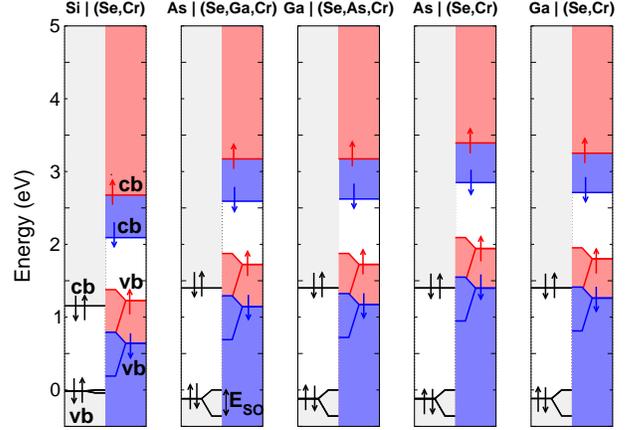}}
\caption{ \label{offsets1_fig} Valence and conduction band offsets for heterojunctions with 
Si and GaAs for majority and minority spin channels. Spin-orbit splitting of the
valence bands is denoted. The notation labeling the interfaces is described
in the caption for Table II.
}
\end{figure}

Our calculated valence- and conduction-band offsets for the
five interfaces described in Sec.~\ref{InterfaceModels} are summarized
graphically in Fig.~\ref{offsets1_fig} and tabulated numerically in
Table~\ref{tab_offsets}. In Table~\ref{tab_offsets} we show only offsets
for the valence bands since only these were evaluated completely from first 
principles. For all five interfaces studied, we find that the
\cdcrse\ conduction-band edges for both spin channels 
sit well above the conduction-band edge of the substrate---independent
of the substrate material (Si or GaAs), termination layer (Ga or As),
and the atomic intermixing described in Sec.~\ref{InterfaceModels}.
The band alignments are all Type IIB for the majority spin channel,
and either IIA or weakly IIB for the minority channel.
Thus, according to the criteria outlined in
Sec.~\ref{Introduction}, these band alignments are not favorable for
efficient spin injection.

A major goal of this work is to identify specific interface features
that strongly affect band offsets, and to evaluate whether such
features can be externally controlled, for example, during the growth
process. For several of the interfaces we have studied, we have
identified one such feature: the concentration of Cr atoms at the
interface.  Our motivation for studying the effects of varying the
interfacial Cr content comes from analyzing the environmental
dependence of Cr removal energies. For example, the calculated
formation energy for a Cr vacancy in bulk
\cdcrse\ is 3.27 eV, but for Cr atoms in the interface layer of a
\cdcrse/Si interface the formation of a single vacancy is actually
exothermic, with an energy gain of 1.60 eV/vacancy.\cite{note}

To explore the impact of interfacial Cr content on band offsets, we
consider the limiting case of total depletion of Cr from the mixed (Se,Cr) 
layer at the interface.  Since the Cr atoms do not
participate directly in the tetrahedral bonding across the interface
boundary, their removal leads to relatively minor rearrangements of
the interface structure---which we nevertheless fully account for by
re-optimizing the geometry.  The energetics of
removing this Cr layer vary somewhat from interface to interface, but
even in the limiting case of a Cr-rich growth environment, the total
depletion of Cr from the final interface layer is thermodynamically
favorable for all five interfaces, as shown in Table~\ref{form_energy_tab}.

\begin{figure}[bt]
\centerline{\epsfig{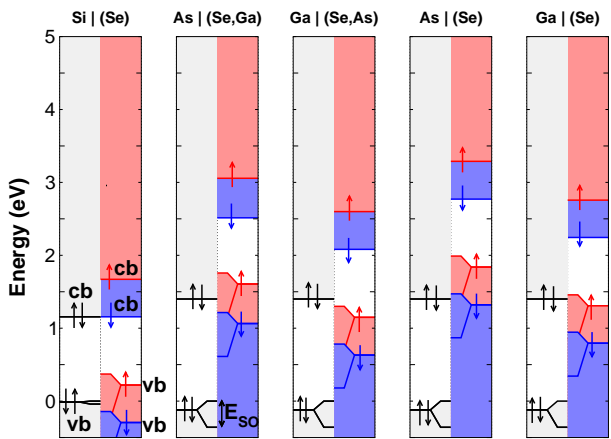}}
\caption{ \label{offsets2_fig} Valence and conduction band offsets for heterojunctions with 
Si and GaAs, in which the interface layers has been depleted of Cr.
}
\end{figure}

The band alignments and offsets for Cr-depleted interfaces are
summarized in Fig.~\ref{offsets2_fig}. Compared to the non-depleted
interfaces, all of the offsets are reduced. The reduction is
remarkably large for the interface with Si (about 1 eV), smaller for
both abrupt and intermixed Ga-terminated GaAs interfaces (about
0.5 eV), and smallest for As-terminated GaAs interfaces (about
0.1 eV). In two cases (Si, and Ga-terminated GaAs) the reduction is
actually sufficient to change the alignment type from IIB to IIA.

Finally, we emphasize that the question of {\it whether} the
concentration of interfacial Cr can be externally controlled remains
open.  Although the energetics clearly favors Cr-depleted interfaces,
kinetic barriers may be more important---and perhaps decisive---at the
relatively low growth temperatures typical of molecular-beam epitaxy.
Theoretical studies of complex interfaces are notoriously difficult,
not the least because of uncertainties in the interface
morphology. The present work clearly demonstrates the possible dangers
of assuming an ``idealized'' interface structure, and should make
clear that an understanding of the systematics of band offsets
requires---at the very least---detailed microscopic calculations.
\begin{table} [tb]
\caption{\label{tab_offsets}
Valence band offsets for heterojunctions of Si and GaAs with CdCr$_2$Se$_4$. 
Offsets are in eV. The notation A$\mid$(B,C) denotes a substrate 
terminated by atoms of chemical type A adjacent to interface layers containing 
atoms of type B and C. For example, Fig.~\ref{interface1_fig} is denoted 
Si$\mid$(Se,Cr), whereas Fig.~\ref{interface2_fig} is denoted As$\mid$(Se,Cr,Ga).
}
\begin{tabular}{lcc}
Heterojunction &  VBO$_{\uparrow}$  & VBO$_{\downarrow}$  \\
\hline
\makebox[0.15in][r]{Si}$\mid$(Se,Cr)       & \dec  1.38 &  \dec 0.79 \\
\makebox[0.15in][r]{Si}$\mid$(Se)          & \dec  0.37 &  \dec $-$0.14 \\
\\
\makebox[0.15in][r]{As}$\mid$(Se,Ga,Cr)    & \dec  1.87 &  \dec 1.29 \\  
\makebox[0.15in][r]{As}$\mid$(Se,Ga)       & \dec  1.78 &  \dec 1.21 \\ 
\makebox[0.15in][r]{As}$\mid$(Se,Cr)       & \dec  2.09 &  \dec 1.55 \\        
\makebox[0.15in][r]{As}$\mid$(Se)          & \dec  1.99 &  \dec 1.47 \\ 
\\
\makebox[0.15in][r]{Ga}$\mid$(Se,As,Cr)    &  \dec 1.87 &  \dec 1.32  \\     
\makebox[0.15in][r]{Ga}$\mid$(Se,As)       &  \dec 1.30 &  \dec 0.78 \\   
\makebox[0.15in][r]{Ga}$\mid$(Se,Cr)       &  \dec 1.95 &  \dec 1.41 \\    
\makebox[0.15in][r]{Ga}$\mid$(Se)          &  \dec 1.46 &  \dec 0.94 \\ 
\end{tabular}
\end{table}
 TABLE
\begin{table} [tb]
\caption{\label{form_energy_tab}
Energy gained by depleting Cr from the interface layer, in eV per vacancy.
\cite{note} The labels refer to the parent heterojunction before depletion of Cr.
}
\begin{tabular}{lc}
Heterojunction & $\Delta E$ \\
\hline
\makebox[0.15in][r]{Si}$\mid$(Se,Cr)    & \dec 0.69 \\
\makebox[0.15in][r]{As}$\mid$(Se,Ga,Cr) & \dec 0.81 \\
\makebox[0.15in][r]{As}$\mid$(Se,Cr)    & \dec 0.38 \\
\makebox[0.15in][r]{Ga}$\mid$(Se,As,Cr) & \dec 0.12 \\
\makebox[0.15in][r]{Ga}$\mid$(Se,Cr)    & \dec 0.91 \\
\end{tabular}
\end{table}


\section{Summary}\label{Summary}
In summary we have applied first principles methods to determine the absolute
valence band offsets for heterojunctions between \cdcrse\ an {\it n}-type ferromagnetic
semiconductor, and Si or GaAs. By using experimentally determined band gaps, we have also
determined the type of band alignments in these heterojunctions, and find
favorable alignment for electrical injection for Cr-depleted heterojunctions 
with Si and with Ga-terminated GaAs substrates. These findings suggest that 
\cdcrse\ may be a promising candidate for the injection of spin-polarized electrons.
Finally, we point out that although the type Type IIB alignments we have found are not
favorable for spin injection, such alignments are ideal for spin-polarized resonant
interband tunneling diodes in which \cdcrse\ would function as a spin-selective
quantum well.

\section{Acknowledgements}
One of the authors (J.M.S.) acknowledges the National Research
Council for support during this work in the form of a postdoctoral fellowship. 
We thank Andre Petukhov, Steve Hellberg, Berry Jonker and Aubrey Hanbicki
for many useful conversations during the course of this work. This work was 
funded in part by DARPA and ONR. Computational work was supported in part by a grant of 
HPC time from the DoD Major Shared Resource Center ASCWP.


\begin{references}
\bibitem{note} To evaluate these formation energies we have used the energy per atom of bulk
Cr as the chemical potential of Cr. Although the conditions of the
growth process may lead to a different numerical value the bulk energy per atom is the
thermodynamic upper bound, and hence any lower value leads to a larger energy gain for Cr depletion.
\end{references}
\end{document}